# Influences of spin-orbit coupling on Fermi surfaces and Dirac cones in ferroelectric-like polar metals

Hu Zhang, Wei Huang, Jia-Wei Mei, and Xing-Qiang Shi*

Department of Physics, Southern University of Science and Technology, Shenzhen 518055, China

Based on first-principles calculations and $k \cdot p$ effective models, we report physical properties of ferroelectric-like hexagonal polar metals with $P6_3mc$ symmetry, which are distinct from those of conventional metals with spatial inversion symmetry. Spin textures exist on the Fermi surface of polar metals (e.g. ternary LiGaGe and elemental polar metal of Bi) and spin-momentum locking exist on accidental Dirac cones on the sixfold rotational axis, due to the spin-orbit coupling and lack of inversion symmetry in polar space group. This effect has potential applications in spin-orbitronics. Dirac points are also predicted in LiGaGe.

## I. INTRODUCTION

Based on the existence of spatial inversion symmetry, metals can be divided into two classes: centrosymmetric and non-centrosymmetric metals. Most of well-known metals are centrosymmetric. Non-centrosymmetric metals can be further divided into non-polar metals (e.g. CoSi with $P2_13$ symmetry [1]) and polar metals, as classified in Fig. 1. In 1965, Anderson and Blount proposed the existence of metals with polar space groups in which the inversion symmetry is breaking [2]. In 2013, the first experimental evidence of a ferroelectric-like structural transition was reported in metallic $LiOsO_3$ at 140 K [3]. After that, the origin of this transition was extensively investigated [4-8]. Some other ferroelectric-like polar metals have also been discovered [9-12], and single-element ferroelectric-like polar metals have been proposed [13].

In this work, we will study two previously unrecognized physical phenomena in (nonmagnetic) polar metals with the consideration of spin-orbit coupling (SOC) effects: spin textures on the Fermi surface and spin-momentum locking on Dirac cones. For metals with both spatial inversion symmetry and time-reversal symmetry, all the energy bands are at least doubly-degenerate. Due to the lack of inversion symmetry in polar metals, SOC splits the level degeneracies of energy bands except for the bands at high symmetry points (e.g. the Brillouin zone center) and along high symmetry directions. This could result in spin textures on the Fermi surface of polar metals, which is absent in centrosymmetric (nonmagnetic) metals such as Li and Au. Understanding of this previously unexplored property is of interest for practical applications of ferroelectric-like polar metals. On the other hand, accidental Dirac points with fourfold degeneracy can exist at the high rotational axis of polar metals, while SOC lifts the band degeneracy around Dirac points along other symmetry directions. Such Dirac fermions have been predicted in hexagonal polar metals LiZnBi [14], CaAgBi [15], and SrHgPb [16] recently. Here we will show that there exists spin-momentum locking on Dirac cones in polar metals.

Previous works identified a class of hexagonal ferroelectrics with the LiGaGe-type structure (space group $P6_3mc$) [17]. Here we show that LiGaGe, which has been synthesized experimentally, is actually a polar metal and has Dirac points. We take LiGaGe as an example to study spin textures on the Fermi surface and spin-momentum locking on Dirac cones in polar metals using the $k \cdot p$ effective models and first-principles calculations with SOC. Similar spin textures are also studied here in recently predicted ferroelectric-like *elemental* polar metals ($P6_3mc$) formed by group-V elements [13]. Then, we classify crystal symmetry protected accidental Dirac points in polar metals with the $P6_3mc$ symmetry by analyzing compatibility relations between the irreducible representations of energy bands along the high rotational axis with the consideration of SOC. Finally, we investigate spin-momentum locking on Dirac cones in polar metals.

## II. METHODOLOGY

To investigate physical properties of ferroelectric-like polar metals, $k \cdot p$ effective models combined with density functional theory (DFT) [18] calculations were performed. In first-principles DFT calculations, we adopt the Vienna Ab Initio Simulation Package (VASP) [19-21] with the Perdew–Burke–Ernzerhoff (PBE) functional in generalized gradient approximation (GGA) [22]. An energy cutoff of 500 eV was used for all calculations. The metallic behavior of LiGaGe was checked with the HSE06 hybrid function [23]. The $k \cdot p$ low energy effective Hamiltonian was constructed based on the theory of invariants [24].

## III. RESULTS AND DISCUSSION

### A. spin textures on Fermi surface

In LiGaGe with the $P6_3mc$ symmetry, Ga and Ge atoms form buckling planes separated with the stuffing Li atoms, as shown in Fig. 2(a). Our theoretical lattice constants are $a$ = 4.23 Å, $c$ = 6.88 Å, which consist with experimental values ($a$ = 4.18 Å, $c$ = 6.78 Å) [17]. Figure 2(b) gives the Brillouin zone with high symmetry points indicated for crystals with the $P6_3mc$ symmetry. The electronic band structures of LiGaGe calculated with SOC are plotted in Fig. 2(c). The energy bands near the Γ (the Brillouin zone center) and M points cross the Fermi level. Thus, LiGaGe is a ferroelectric-like polar metal. We can also find band crossing along the Γ-A direction (the sixfold rotational $k_z$ axis).

Based on symmetry analyses, one can capture general characters of energy bands near the Γ point. For the Brillouin zone shown in Fig. 2(b), the point group of wave vector $k$ at Γ is $C_{6v}$. The point group $C_{6v}$ consists of two generators including a mirror symmetry $\sigma_v$ ($x$, $-y$, $z$) and a sixfold rotation $c_{6z}$ around the $z$ axis. States at Γ transform according to the $\Gamma_7$, $\Gamma_8$, and $\Gamma_9$ double-group irreducible representations of $C_{6v}$ taking into account SOC. These three irreducible representations are all two-dimensional. Considering the constraint of the low-energy effective model $H(k)$ with respect to crystal and time-reversal symmetry [25], $H(k)$ for the $\Gamma_7$ band around the Γ point up to third order in $k$ is given by

$$H(k) = E_0(k) + E_1(k)(k_x\sigma_y - k_y\sigma_x), \tag{1}$$

where $E_0(k) = c_1 + c_2(k_x^2 + k_y^2) + c_3 k_z^2$, $E_1(k) = c_4 + c_5(k_x^2 + k_y^2) + c_6 k_z^2$, and $\sigma_{x,y}$ are the Pauli matrices. The main character of the $\Gamma_7$ band is the appearance of the linear term $c_4(k_x\sigma_y - k_y\sigma_x)$, which is known as the Rashba term [26]. The corresponding band dispersion is

$$E_\pm(k) = E_0(k) \pm E_1(k)\sqrt{k_x^2 + k_y^2}. \tag{2}$$

In Fig. 3(a), we plot the $\Gamma_7$ band along $k_x$. The SOC induced linear term in $k$ around the $\Gamma$ point is obvious. The low-energy effective model of the $\Gamma_8$ band is the same as that of the $\Gamma_7$ band.

For the $\Gamma_9$ band, the Hamiltonian around $\Gamma$ up to third order in $k$ takes the from

$$H(k) = E_0(k) + A_1(k)\sigma_x + A_2(k)\sigma_y, \tag{3}$$

where $E_0(k) = c_1 + c_2(k_x^2 + k_y^2) + c_3 k_z^2$, $A_1(k) = c_4(3k_x^2 k_y - k_y^3)$, $A_2(k) = c_5(k_x^3 - 3k_x k_y^2)$. The band dispersion is

$$E_\pm(k) = E_0(k) \pm \sqrt{A_1(k)^2 + A_2(k)^2}. \tag{4}$$

The $\Gamma_9$ band along $k_x$ is plotted in Fig. 3(b), which shows a very different character compared with the $\Gamma_7$ band. This is due to the absence of the linear term in $k$ for the $\Gamma_9$ band.

We now investigate physical properties of the Fermi surface in real materials. Electronic band structures of LiGaGe along M-$\Gamma$-K directions are shown in Fig. 4(a). The irreducible representations of bands at the $\Gamma$ point are also indicated. The doubly-degenerate $\Gamma_7$ and $\Gamma_9$ states lie just below and above the Fermi level, respectively. Away from the $\Gamma$ point, the band degeneracy is lifted. As a result, the $\Gamma_7$ band exhibits linear dispersions. On the contrary, the $\Gamma_9$ band exhibits non-linear dispersions. These characters consist with above general discussions based on the model Hamiltonian. Furthermore, the spin split $\Gamma_7$ band will cross the Fermi level. In Fig. 4(b), we show the Fermi surface in the M$\Gamma$K ($k_z = 0$) plane. The Fermi surface is sixfold symmetric, reflecting the crystal symmetry of LiGaGe. Open lines come from bands near the M point (originating from the $\Gamma_9$ state at the $\Gamma$ point) as can be confirmed in Fig. 2(c). The remaining part comes from dispersions of the $\Gamma_7$ band. The spin-textures on the Fermi surface are shown in Fig. 4(c). For usual centrosymmetric metals (nonmagnetic, such as Li and Au), the energy bands at each general momentum *k* are doubly-degenerate due to the coexistence of spatial inversion symmetry and time-reversal symmetry. As a result, there are no net spin-textures on the Fermi surface of centrosymmetric metals. From Fig. 4(c) we can find that the spin is not perpendicular to the momentum generally. This indicates the complexity of spin-textures in real materials for the large momentum *k*.

Next we take bismuth (Bi) as an example to consider SOC effects in ferroelectric-like elemental polar metals formed by group-V elements [13]. The PBE theoretical lattice constants are *a* = 4.45 Å, *c* = 8.88 Å, in qualitative accordance with previous local density approximation results [13]. The electronic band structures shown in Fig. 5(a) display metallic behavior of polar phase Bi. At the Γ point, two doubly-degenerate $\Gamma_9$ states lie near the Fermi level. Away from the Γ point, one of the spin split $\Gamma_9$ bands cross the Fermi level. There are also two other bands crossing the Fermi level along the Γ-M direction. These bands originate from the $\Gamma_7$ state at the Γ point. The Fermi surface and corresponding spin textures in the MΓK ($k_z = 0$) plane are shown in Fig. 5(c). Around the Γ point (inner bands), the spin is always perpendicular to the momentum. This is a typical spin-momentum locking effect. More complex spin textures can be found for other bands.

### B. spin-momentum locking on Dirac cones

Another interesting phenomenon in ferroelectric-like polar metals with $P6_3mc$ symmetry is the emergence of symmetry protected accidental Dirac points. In the Brillouin zone shown in Fig. 2(b), points along Γ-A (the $k_z$ axis) can be denoted as Δ (0, 0, u). Both point groups of wave vector *k* at Δ and A (0, 0, 0.5) are all $C_{6v}$ [27]. Similar to the case of states at the Γ point, states at Δ can only belong to three two-dimensional $\Delta_7$, $\Delta_8$, and $\Delta_9$ irreducible representations of $C_{6v}$ taking into account SOC. The band crossing is only allowed between bands with different irreducible representations. Therefore, accidental Dirac points on the $k_z$ axis may be originated from the band crossing between: $\Delta_7$ and $\Delta_8$ bands, $\Delta_7$ and $\Delta_9$ bands, and $\Delta_8$ and $\Delta_9$ bands. The situation is very different at the Brillouin zone boundary point A (0, 0, 0.5) due to the nonsymmorphic nature of the space group $P6_3mc$. Bands will stick together at A. There are two two-dimensional irreducible representations $A_4$ and $A_5$, and one four-dimensional irreducible representation $A_6$ for the A point. In Table I, we list compatibility relations between the irreducible representations of the group of the wave vector along the Γ-Δ-A direction. It should be noted that Δ point bands stick together in pairs at the A point. Based on these results, we can classify different accidental Dirac points on the $k_z$ axis. Figure 6(a) shows a Dirac point formed with the band crossing between $\Delta_7$ and $\Delta_8$ bands. This kind of Dirac point is usually called the type-I since slopes of the two crossing bands have opposite signs [28]. In Fig. 6(b), there are two associated Dirac points. One is a type-I Dirac point coming from the band crossing between $\Delta_7$ and $\Delta_9$ bands. Interestingly, the slopes of the two crossing $\Delta_8$ and $\Delta_9$ bands have the same sign. This results in a type-II Dirac point [28,29]. Such tilted Dirac cone can have very unusual Landau level spectrum compared to that of type-I Dirac fermions [30]. Materials with type-II Dirac points may have special magnetotransport properties. The coexistence of type-II and type-I Dirac points on the sixfold rotational axis originates from the sticking of $\Delta_7$ and $\Delta_8$ bands, which is a direct result of the nonsymmorphic crystal structures. With the shift of the energy of $A_{4,5}$ bands, we can also obtain a type-II Dirac point formed with $\Delta_7$ and $\Delta_9$ bands and a type-I Dirac point formed with $\Delta_8$ and $\Delta_9$ bands.

We now study spin-momentum locking around Dirac points in polar metals. Figure 7(a) shows electronic band structures of CaAgAs with the LiGaGe-type structure, which was predicted to be a Dirac semimetal previously [15], along the Γ-A direction calculated with SOC. At the Γ point, doubly-degenerate $Γ_9$ and $Γ_7$ states lie above and below the Fermi level respectively. Along the Γ-A direction, we can find a type-I Dirac point due to the band crossing of $\Delta_7$ and $\Delta_9$ bands. The Dirac point is located at the D (0, 0, $u$ = 0.0675) (in units of $2π/c$) point. In Fig. 7(b), we plot electronic band structures along the (0.5, 0.0, u)-D-(1/3, 1/3, $u$) directions (in the $k_z$ = $u$ plane). Due to the lack of inversion symmetry, SOC splits the level degeneracies of the energy bands. Figure 7(c) shows the Fermi surfaces and spin textures in the $k_z = u$ plane at 0.05 eV. The outer and inner bands have clockwise and counterclockwise spin textures, respectively. The spin is always perpendicular to the momentum, which reveals a spin-momentum locking around Dirac points in ferroelectric-like polar metals. To further understand $\Gamma_7$ and $\Gamma_9$ states, we can construct a low-energy effective Hamiltonian $H(k)$ around the $\Gamma$ point [24]. We find that $H(k)$ is given by

$$H(k) = \epsilon_0(k) + \begin{pmatrix} M(k) & -iDk_- & B(k)k_+ & -Ak_-^2 \\ iDk_+ & M(k) & Ak_+^2 & B(k)k_- \\ B^*(k)k_- & Ak_-^2 & -M(k) & 0 \\ -Ak_+^2 & B^*(k)k_+ & 0 & -M(k) \end{pmatrix}, \quad (5)$$

where $k_\pm = k_x \pm ik_y$, $\epsilon_0(k) = C_0 + C_1 k_z^2 + C_2(k_x^2 + k_y^2)$, $M(k) = M_0 - M_1 k_z^2 - M_2(k_x^2 + k_y^2)$, $B(k) = -iB_1 + B_2 k_z$, and all the model parameters are real. For momentum $k_z$ along the $\Gamma$-A direction ($k_x = k_y = 0$), we have

$$M(k_z) = M_0 - M_1 k_z^2, \quad (6)$$

where $M_0 M_1 > 0$ describes a band-inverted state. In this case we get two Dirac points (crossing points) at $k_z^c = \pm\sqrt{M_0/M_1}$. We can expand the Hamiltonian near the Dirac point up to linear order in $q$, by defining $q = k - k_z^c = (q_x, q_y, q_z)$ and choosing $k_z^c$ for reference. We then find that

$$H(q) = (2C_1 k_z^c)q_z + \begin{pmatrix} -(2M_1 k_z^c)q_z & -iDq_- & -iB_1 q_+ & 0 \\ iDq_+ & -(2M_1 k_z^c)q_z & 0 & -iB_1 q_- \\ iB_1 q_- & 0 & (2M_1 k_z^c)q_z & 0 \\ 0 & iB_1 q_+ & 0 & (2M_1 k_z^c)q_z \end{pmatrix}. \quad (7)$$

The band dispersion for momentum $q_z$ is $E_\pm(q_z) = 2k_z^c(C_1 \pm M_1)q_z$. On the other hand, the band dispersion for momentum in the $q_z = 0$ plane is

$$E(q_{x,y}) = v_k \sqrt{q_x^2 + q_y^2}, \quad (8)$$

with

$$v_k = \pm\sqrt{B_1^2 + \frac{1}{2}D^2 \pm \frac{1}{2}D\sqrt{4B_1^2 + D^2}}. \quad (9)$$

The in-plane velocity around the Dirac points has two different values as a result of inversion symmetry breaking.

Figure 8 shows electronic band structures of LiGaGe along the Γ-A direction calculated with SOC. There exist accidental Dirac points just below the Fermi level. We find that Dirac points formed with $\Delta_8$ and $\Delta_9$ bands are also described by the effective model given in (7). For $\Gamma_7$ and $\Gamma_8$ states (see Appendix), we find that the Hamiltonian near the Dirac point up to linear order in $q$ is given by

$$H(q) = \begin{pmatrix} h_1(q) & 0 \\ 0 & h_2(q) \end{pmatrix}, \tag{10}$$

with $h_1(q) = 2(C_1 - M_1)k_z^c q_z + (A_1 + A_2)(q_x \sigma_y - q_y \sigma_x)$, $h_2(q) = 2(C_1 + M_1)k_z^c q_z + (A_1 - A_2)(q_x \sigma_y - q_y \sigma_x)$. The band dispersion for momentum $q_z$ is $E_\pm(q_z) = 2k_z^c(C_1 \pm M_1)q_z$. For momentum in the $q_z = 0$ plane we have

$$E(q_{x,y}) = \pm(A_1 \pm A_2)\sqrt{q_x^2 + q_y^2}. \tag{11}$$

Again, the in-plane velocity around the Dirac points has two different values $(A_1 \pm A_2)$ as a result of inversion symmetry breaking. The Hamiltonian (11) also produces spin-momentum locking around Dirac points similar to the case in Fig. 7(c).

### C. outlooks

Due to the unique properties of Fermi surface and Dirac points in ferroelectric-like polar metals, e.g. LiGaGe, spin-dependent transport phenomena in polar metals are promising avenues for further studies. There may exist unusual magnetoconductivity in LiGaGe at magnetic field [31]. Spin textures on the Fermi surface and spin-momentum locking on accidental Dirac cones in polar metals can be observed by means of spin- and angle-resolved photoelectron spectroscopy [32]. Finally, the spin-momentum locking around Dirac points also makes such materials having potential applications in Datta-Das spin transistors [26,33].

### IV. CONCLUSIONS

In summary, we have investigated influences of spin-orbit coupling on Fermi surface and Dirac cones in ferroelectric-like hexagonal polar metals with the *P*6$_3$*mc* symmetry based on first-principles calculations and the *k · p* methods. Taking LiGaGe and Bi as examples, we have studied spin-textures on the Fermi surface. On the other hand, accidental Dirac points on the sixfold rotation axis are classified with the help of compatibility relations between the irreducible representations of the group of the wave vector. There exists spin-momentum locking on Dirac cones in polar metals. We also find Dirac points in LiGaGe.

## ACKNOWLEDGMENTS

This work was supported by the Shenzhen Fundamental Research Foundation (Grant No. JCYJ20170817105007999), the Natural Science Foundation of Guangdong Province of China (Grant No. 2017A030310661) and the Natural Science Foundation of China (Grant No. 11474145).

## APPENDIX: EFFECTIVE HAMILTONIAN

We can construct a low-energy effective Hamiltonian $H(k)$ around the $\Gamma$ point to understand $\Gamma_7$ and $\Gamma_8$ states (see also Fig. 6(a)) [24]. We find that this Hamiltonian $H(k)$ is given by

$$H(k) = \epsilon_0(k) + \begin{pmatrix} M(k) & -i(A_1+A_2)k_- & 0 & -iBk_+^2 \\ i(A_1+A_2)k_+ & M(k) & -iBk_-^2 & 0 \\ 0 & iBk_+^2 & -M(k) & -i(A_1-A_2)k_- \\ iBk_-^2 & 0 & i(A_1-A_2)k_+ & -M(k) \end{pmatrix}, \quad (1)$$

where $k_\pm = k_x \pm ik_y$, $\epsilon_0(k) = C_0 + C_1 k_z^2 + C_2(k_x^2+k_y^2)$, $M(k) = M_0 - M_1 k_z^2 - M_2(k_x^2+k_y^2)$, and all the model parameters are real. We can get two Dirac points at $k_z^c = \pm\sqrt{M_0/M_1}$ when $M_0 M_1 > 0$. This Hamiltonian can also be expanded by defining $q = k - k_z^c = (q_x, q_y, q_z)$ and choosing $k_z^c$ for reference. The Hamiltonian up to linear order in $q$ is given by

$$H(q) = \begin{pmatrix} h_1(q) & 0 \\ 0 & h_2(q) \end{pmatrix}, \quad (2)$$

with $h_1(q) = 2(C_1 - M_1)k_z^c q_z + (A_1+A_2)(q_x\sigma_y - q_y\sigma_x)$, $h_2(q) = 2(C_1 + M_1)k_z^c q_z + (A_1-A_2)(q_x\sigma_y - q_y\sigma_x)$.

TABLE I. Compatibility relations between the irreducible representations of the group of the wave vector $k$ along $\Gamma(0, 0, 0)$ - $\Delta(0, 0, u)$-$A(0, 0, 0.5)$ [27].

| $\Gamma$ point representation | | $\Delta$ point representation | $\Delta$ point representation | | A point representation |
|---|---|---|---|---|---|
| $\Gamma_7$ | $\rightarrow$ | $\Delta_7$ | $\Delta_7 + \Delta_8$ | $\rightarrow$ | $A_6$ |
| $\Gamma_8$ | $\rightarrow$ | $\Delta_8$ | $2\Delta_9$ | $\rightarrow$ | $A_4 + A_5$ |
| $\Gamma_9$ | $\rightarrow$ | $\Delta_9$ | | | |

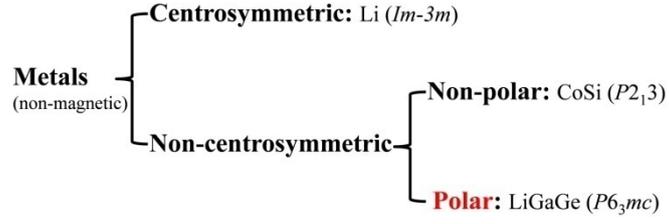

FIG. 1. Classifications of metals based on space group symmetry. Examples are given.

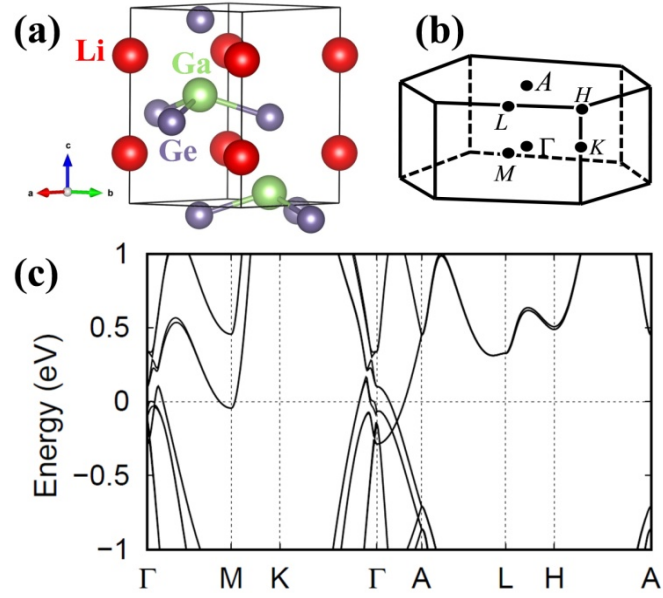

FIG. 2. (a) The crystal structures of LiGaGe ($P6_3mc$). (b) Brillouin zone of LiGaGe. Points between Γ (0, 0, 0) and A (0, 0, 0.5) are denoted as Δ (0, 0, u) in Table I and Figure 6 below. (c) Electronic band structures of LiGaGe calculated with SOC. The Fermi level is set at 0 eV.

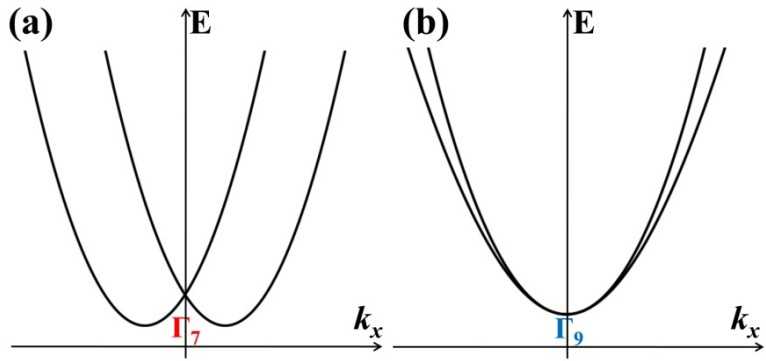

FIG. 3. The (a) $\Gamma_7$ and (b) $\Gamma_9$ bands along $k_x$.

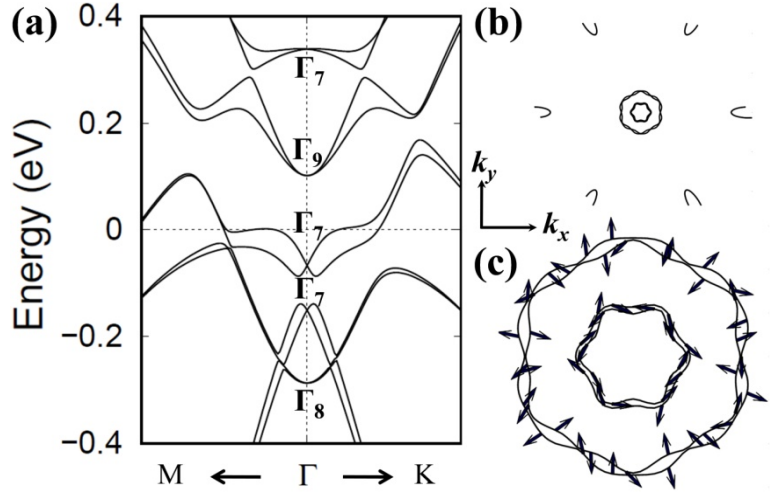

FIG. 4. (a) Electronic band structures of LiGaGe along M-Γ-K directions calculated with SOC. The Fermi level is set at 0 eV. (b) Fermi surface of LiGaGe in the MΓK ($k_z = 0$) plane and (c) spin textures around the Γ point.

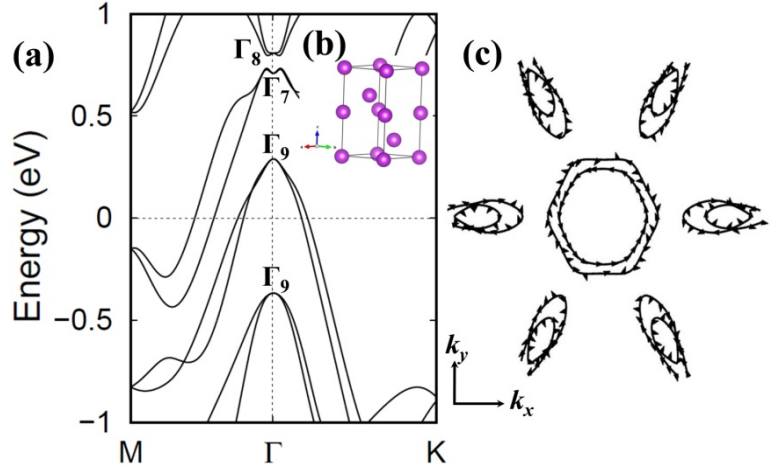

FIG. 5. (a) Electronic band structures of Bi with a polar space group $P6_3mc$ along M-Γ-K directions calculated with SOC. The Fermi level is set at 0 eV. (b) The crystal structures of the polar phase Bi in a distorted α-La-like structure ($P6_3mc$) in which two inequivalent Wyckoff positions 2a (0, 0, 0) and 2b (1/3, 2/3, 0.1923) are occupied. (c) Fermi surface and spin textures of the polar phase Bi in the MΓK ($k_z = 0$) plane.

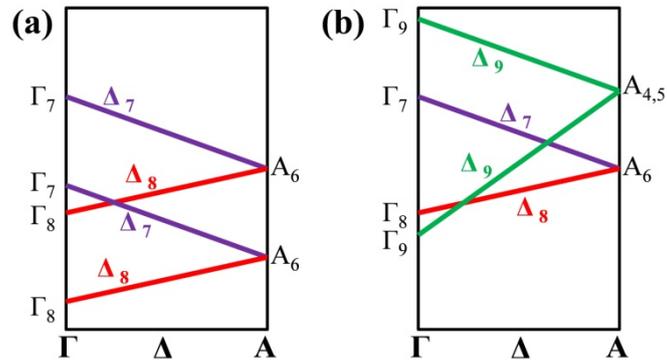

FIG. 6. The classifications of accidental Dirac points on the sixfold rotational axis of crystals with $P6_3mc$ symmetry based on compatibility relations shown in Table I.

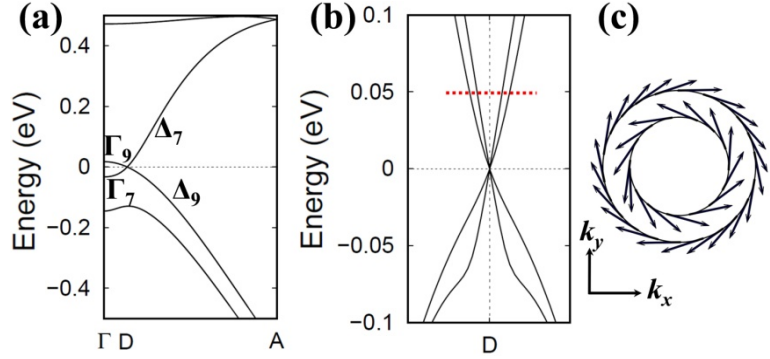

FIG. 7. (a) Electronic band structures of CaAgAs along the Γ-A direction calculated with SOC. The Dirac point is located at the D (0, 0, u = 0.0675) (in units of $2\pi/c$) point. (b) Electronic band structures along (0.5, 0.0, u)-D-(1/3, 1/3, u) directions. (c) Fermi surface and spin textures in the $k_z = u$ plane at 0.05 eV [dashed line in (b)].

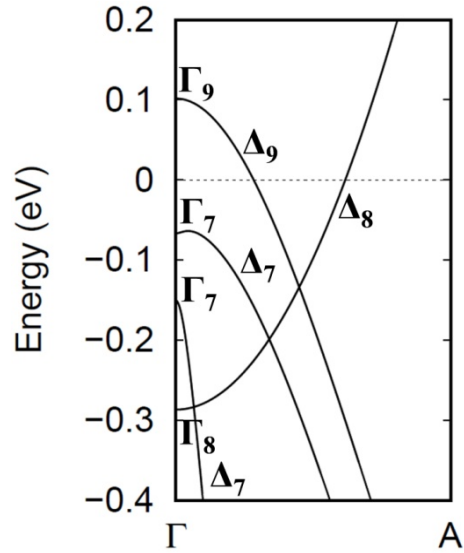

FIG. 8. (a) Electronic band structures of LiGaGe along the Γ-A direction calculated with SOC.

* shixq@sustc.edu.cn